\documentclass{article}
\usepackage{emulateapj,pstricks,apjfonts}

\def\axaf       {{\em AXAF}\/}
\def\asca       {{\em ASCA}\/}
\def\einstein   {{\em Einstein}\/}
\def\rosat      {{\em ROSAT}\/}
\def\kmsmpc     {~km$\;$s$^{-1}\,$Mpc$^{-1}$}
\def\ergs       {~erg$\;$s$^{-1}$}
\def\ergcms     {~erg$\;$s$^{-1}\,$cm$^{-2}$}
\def\gax        {\gtrsim}
\def\lax        {\lesssim}
\def\msun       {~$M_{\odot}$}

\begin{document}

\submitted{ApJ in press, 1998 September 1}

\lefthead{THE $L_X-T$ RELATION AND TEMPERATURE FUNCTION}
\righthead{MARKEVITCH}

\title{THE $L_X-T$ RELATION AND TEMPERATURE FUNCTION FOR NEARBY CLUSTERS
REVISITED} 

\author{Maxim Markevitch\altaffilmark{1}}

\affil{Harvard-Smithsonian Center for Astrophysics, 60 Garden St.,
Cambridge, MA 02138; maxim@head-cfa.harvard.edu}

\altaffiltext{1}{Also Space Research Institute, Russian Academy of Sciences}

\begin{abstract}

The X-ray luminosity-temperature relation for nearby $T\simeq 3.5-10$ keV
clusters is rederived using new \asca\ temperatures (Markevitch et al.\
1998a) and \rosat\ luminosities. Both quantities are derived by directly
excluding the cooling flow regions. This correction results in a greatly
reduced scatter in the $L_X-T$ relation; cooling flow clusters are similar
to others outside the small cooling flow regions. For a fit of the form
$L_{\rm bol}\propto T^\alpha$, we obtain $\alpha=2.64\pm0.27$ (90\%) and a
residual rms scatter in $\log L_{\rm bol}$ of 0.10. The derived relation can
be directly compared to theoretical predictions that do not include
radiative cooling.  It also provides an accurate reference point for future
evolution searches and comparison to cooler clusters.

The new temperatures and $L_X-T$ relation together with a newly selected
cluster sample are used to update the temperature function at $z\sim 0.05$.
The resulting function is generally higher and flatter than, although within
the errors of, the previous estimates by Edge et al.\ (1990) and Henry \&
Arnaud (1991, as rederived by Eke et al.\ 1996). For a qualitative estimate
of constraints that the new data place on the density fluctuation spectrum,
we apply the Press-Schechter formalism for $\Omega_0=1$ and 0.3. For
$\Omega_0=1$, assuming cluster isothermality, the temperature function
implies $\sigma_8=0.55\pm0.03$, while taking into account the observed
cluster temperature profiles, $\sigma_8=0.51\pm0.03$, consistent with the
previously derived range. The dependence of $\sigma_8$ on $\Omega_0$ differs
from earlier findings, because of our treatment of the slope of the
fluctuation spectrum $n$ as a free parameter. For the considered values of
$\Omega_0$, $n=-(2.0-2.3)\pm0.3$, somewhat steeper than derived from the
earlier temperature function data, in agreement with the local slope of the
galaxy fluctuation spectrum from the APM survey (Baugh \& Gazta\~naga 1996),
and significantly steeper than the standard CDM prediction.

\end{abstract}


\section{INTRODUCTION}

Clusters exhibit a correlation between their X-ray luminosity and
temperature, approximately $L_{\rm bol}\propto T^3$ (Mushotzky 1984; Edge \&
Stewart 1991; David et al.\ 1993; Fukazawa 1997). However, models of cluster
evolution that assume no segregation of dark and gaseous matter predict
$L_{\rm bol}\propto T^2$ (e.g., Kaiser 1986; Navarro, Frenk, \& White 1995).
This discrepancy suggests that the gas is distributed and evolves
differently from the dark matter. One of the proposed reasons for this
difference is preheating of the intracluster gas by an agent other than
gravity, such as supernovae, at some high redshift (Kaiser 1991; Evrard \&
Henry 1991). Significant energy injection to the gas should have been made
at the time of its enrichment with heavy elements (e.g., David, Forman, \&
Jones 1991; Loewenstein \& Mushotzky 1996), although neither the amount nor
the epoch of such injection is certain. Models that allow preheating succeed
in approximately reproducing the observed $L_X-T$ slope (e.g., Navarro et
al.\ 1995). If the gas was preheated and later subjected to merger shocks,
cooler clusters and groups are expected to have a steeper $L_X-T$ dependence
than hot clusters (e.g., Cavaliere et al.\ 1997), for which there is some
observational indication (Ponman et al.\ 1996). Also, preheating should make
the redshift evolution of the $L_X-T$ relation slower or absent (Evrard
\& Henry 1991). Indeed, observations exclude any dramatic evolution out to
$z\simeq 0.3-0.5$ (Tsuru et al.\ 1996; Mushotzky \& Scharf 1997), although
this may also be expected in low $\Omega_0$ models without preheating (e.g.,
White \& Rees 1978; Eke, Navarro, \& Frenk 1998).

The previously reported $L_X-T$ relation at both low and high redshifts
exhibits a large scatter which precludes accurate determination of its exact
shape and evolution. Fabian et al.\ (1994) noted that the scatter is mostly
due to the strong cooling flow clusters. Cooling flow regions, present in
more than a half of all low-redshift clusters (e.g., Edge, Stewart, \&
Fabian 1992), are governed by a different physics than the bulk of the
cluster gas, while in extreme cases emitting most of the X-ray luminosity
and strongly biasing the wide-beam temperature measurements (e.g., Allen
1998). Cluster evolution models mentioned above predict global properties of
the cluster gas and generally ignore the presence of cooling flows, thus it
is not obvious that their predictions can be meaningfully compared to the
observations. In principle, radiative cooling can be included in the models,
but it appears that with greater certainty its effects can be separated in
the data, which we attempt in this paper.

The $L_X-T$ relation is a necessary tool for derivation of the cluster
temperature function using a flux-limited sample. For low-redshift clusters,
this function was previously obtained by Edge et al.\ (1990, hereafter E90)
and Henry \& Arnaud (1991, hereafter HA; see correction in Eke, Cole, \&
Frenk 1996). Since the gas temperature is linked to the total cluster mass,
the temperature function provides information on the spectrum of the
cosmological density fluctuations (e.g., HA), as well as on $\Omega_0$ with
additional data from higher $z$ (e.g., Henry 1997 and references therein).
The least certain link in this line of argument is the conversion from the
observed temperature to the cluster mass, which has usually been made
assuming cluster isothermality or near-isothermality. However, recent
\asca\ results (e.g., Markevitch et al.\ 1998a, hereafter M98) indicate
considerable spatial temperature variations which affect the
temperature-mass relation. In addition, the temperature function itself is
affected by cooling flows via sample selection, individual temperature
errors, and the $L_X-T$ relation. All these effects should be corrected for
an accurate comparison with theoretical models.

In this paper, we use the new \asca\ cluster mean temperatures obtained by
directly excluding cooling flow regions (M98) and similarly corrected
luminosities for a cluster sample selected from the new \rosat\ All-Sky
dataset. With these data, we rederive the low-redshift $L_X-T$ relation and
the temperature function which can be directly compared to theoretical
predictions as well as to the oncoming high quality data from higher
redshifts. An accurate comparison with models is best performed by detailed
simulations and is left for future work; for a qualitative estimate of the
constraints that the new data place on the density fluctuation spectrum, we
apply a Press-Schechter formalism for $\Omega_0=1$ and 0.3. We use
$H_0=100\,h$\kmsmpc; confidence intervals are one-parameter 90\% unless
stated otherwise.

\section{THE SAMPLE}

The sample is selected from the \rosat\ All-Sky Survey (RASS) Abell cluster
list (Ebeling et al.\ 1996) plus three known bright non-Abell clusters, with
$0.04\leq z \leq 0.09$ and $f(0.1-2.4\;{\rm keV})> 2\times 10^{-11}$\ergcms.
Because the RASS fluxes have some uncertainty, pointed \rosat\ PSPC or
\einstein\ IPC observations of all clusters with the RASS fluxes greater
than 3/4 of the above limit were analyzed to derive accurate fluxes and
luminosities. The rederived fluxes correspond to the aperture of
$r=1\,h^{-1}$ Mpc (which includes practically all the flux) centered on the
X-ray emission centroid, and exclude all contaminating sources. Below, these
fluxes and corresponding luminosities are referred to as ``total''.

Cooling flows were excised from fluxes and luminosities already at the
sample selection step, for consistency with the analysis below. A
$50\,h^{-1}$ kpc radius contains most of the cooling flow emission in
nonextreme cooling flow clusters such as those in our sample. Therefore, to
do the excision in a uniform manner, for all clusters, regions of
$50\,h^{-1}$ kpc radius centered on the main brightness peak were masked,
and the resulting fluxes and luminosities were multiplied by 1.06 to account
for the flux inside the masked region assuming an average $\beta$-model for
the cluster X-ray brightness, $S_X\propto (1+r^2/a_x^2)^{-3\beta+1/2}$ with
$a_x=125\,h^{-1}$ kpc and $\beta=0.6$ (Jones \& Forman 1984). We refer to
these fluxes and luminosities as ``corrected''. For clusters with strong
cooling flows (e.g., A478, A780, A2597), this correction significantly
reduces the flux; for clusters lacking cooling flows, the flux is (by
design) practically unchanged. Note that this correction procedure is only
approximate and is not indended to result in the exact removal of the
cooling flow contribution, which obviously would require a detailed
modeling. However, its simplicity and model-independence enables a
straightforward comparison to the current simulations that lack resolution
to reproduce cluster high-density regions.

Use of more accurate fluxes from pointed observations has resulted in adding
A3391 and removing A3532 from the original sample, while excision of cooling
flows has resulted in removing A133 and A2597 (with the latter still being
used for the $L_X-T$ relation, as it has been already analyzed). Because
cooling flows only increase the total flux, with high certainty no clusters
from the Ebeling et al.\ (1996) list that satisfy our criteria were missed.
Table~1 lists the final full cluster sample with total and corrected
luminosities and corrected fluxes. Accuracy of the measured fluxes and
luminosities is around 5\% due to the uncertainties in \rosat\ calibration,
background, Galactic absorption, etc.

\noindent
\begin{minipage}{8.9cm}
{\footnotesize
\renewcommand{\arraystretch}{1.2}
\renewcommand{\tabcolsep}{1.5mm}
\begin{center}
TABLE 1
\vspace{1mm}

{\sc Full Cluster Sample}
\vspace{1mm}

\begin{tabular}{lcllrrrr}
\hline \hline
Name  & $z$ & $T_{\rm fit}^{\rm \;a}$ & $T^{\rm \;a}_{\rm corr}$ 
& $L_{X{\rm tot}}^{\rm \;b}$ 
& $L_X^{\rm \;c}$ 
& $L_{\rm bol}^{\rm \;c}$& $f_X^{\rm \;c}$\\
\hline
A85	  &0.052& 6.1 $\pm0.2$& 6.9 $\pm0.4	$   & 2.22& 1.90& 4.54& 6.54 \\
A119	  &0.044& 5.8 $\pm0.6$& 5.6 $\pm0.3	$   & 0.85& 0.88& 1.85& 4.14 \\
A399	  &0.072& 7.4 $\pm0.7$& 7.0 $\pm0.4	$   & 1.66& 1.71& 4.12& 3.07 \\
A401	  &0.074& 8.3 $\pm0.5$& 8.0 $\pm0.4	$   & 2.90& 2.87& 7.49& 4.93 \\
A478	  &0.088& 7.1 $\pm0.4$& 8.4 $^{+0.8}_{-1.4}$& 4.15& 3.21& 8.62& 3.87 \\
A644*	  &0.071& 7.1 $\pm0.6$& 7.9 $\pm0.8     $   & 2.07& 1.87& 4.83& 3.44 \\
A754	  &0.054& 9.0 $\pm0.5$& 9.5 $^{+0.7}_{-0.4}$& 2.19& 2.22& 6.44& 7.07 \\
A780	  &0.057& 3.8 $\pm0.2$& 4.3 $\pm0.4	$   & 1.68& 1.17& 2.12& 3.40 \\
A1644	  &0.048& 4.7 $\pm0.8$ d& 5.0 e             & 0.94& 0.93& ... & 3.79 \\
A1650	  &0.085& 5.6 $\pm0.6$& 6.7 $\pm0.8	$   & 1.88& 1.85& 4.32& 2.39 \\
A1651	  &0.085& 6.3 $\pm0.5$& 6.1 $\pm0.4	$   & 1.97& 1.80& 4.07& 2.35 \\
A1736	  &0.046& 3.5 $\pm0.4$& 3.5 $\pm0.4	$   & 0.50& 0.50& 0.83& 2.18 \\
A1795	  &0.062& 6.0 $\pm0.3$& 7.8 $\pm1.0	$   & 2.60& 1.98& 5.09& 4.78 \\
A2029	  &0.077& 8.7 $\pm0.3$& 9.1 $\pm1.0	$   & 4.28& 3.38& 9.40& 5.39 \\
A2065	  &0.072& 5.4 $\pm0.3$& 5.5 $\pm0.4	$   & 1.29& 1.27& 2.66& 2.24 \\
A2142	  &0.089& 8.8 $\pm0.6$& 9.7 $^{+1.5}_{-1.1}$& 5.22& 4.83&14.19& 5.68 \\
A2256	  &0.058& 7.5 $\pm0.4$& 6.6 $\pm0.4     $   & 2.15& 2.18& 5.06& 6.01 \\
A2319*	  &0.056& 9.2 $\pm0.7$& 8.8 $\pm0.5	$   & 3.67& 3.64&10.06&11.09 \\
A2589	  &0.042& 3.7 ~~~$^{+2.2}_{-1.2}$ d& 3.5 e  & 0.51& 0.44& ... & 2.38 \\
A2597\dag &0.085& 3.6 $\pm0.2$& 4.4 $^{+0.4}_{-0.7}$& 1.73& 0.98& 1.81& 1.26 \\
A2657	  &0.040& 3.7 $\pm0.3$& 3.7 $\pm0.3     $   & 0.47& 0.42& 0.71& 2.44 \\
A3112	  &0.070& 4.7 $\pm0.4$& 5.3 $^{+0.7}_{-1.0}$& 1.64& 1.15& 2.34& 2.16 \\
A3158	  &0.059& 5.5 $\pm0.6$ d& 6.0 e             & 1.41& 1.38& ... & 3.68 \\
A3266	  &0.055& 7.7 $\pm0.8$& 8.0 $\pm0.5	$   & 1.82& 1.83& 4.78& 5.76 \\
A3376	  &0.046& 4.3 $\pm0.6$& 4.0 $\pm0.4	$   & 0.56& 0.57& 0.99& 2.44 \\
A3391	  &0.054& 5.7 $\pm0.7$& 5.4 $\pm0.6     $   & 0.74& 0.74& 1.52& 2.39 \\
A3395	  &0.050& 4.8 $\pm0.4$& 5.0 $\pm0.3	$   & 0.79& 0.81& 1.60& 3.05 \\
A3558	  &0.048& 5.5 $\pm0.3$& 5.5 $\pm0.4	$   & 1.74& 1.68& 3.51& 6.89 \\
A3562	  &0.050& 3.8 $\pm0.9$ d& 4.8 e		    & 0.90& 0.87& ... & 3.21 \\
A3571	  &0.040& 6.9 $\pm0.3$& 6.9 $\pm0.2	$   & 1.97& 1.83& 4.36&10.78 \\
A3667	  &0.053& 7.0 $\pm0.6$& 7.0 $\pm0.6	$   & 2.18& 2.22& 5.34& 7.36 \\
A4059	  &0.048& 4.1 $\pm0.3$& 4.4 $\pm0.3	$   & 0.81& 0.66& 1.22& 2.67 \\
Cyg A*	  &0.057& 6.5 $\pm0.6$& 6.1 $\pm0.4	$   & 2.70& 2.01& 4.46& 5.79 \\
MKW3S	  &0.045& 3.5 $\pm0.2$& 3.7 $\pm0.2     $   & 0.75& 0.54& 0.91& 2.47 \\
Triang*	  &0.051& 9.5 $\pm0.7$& 9.6 $\pm0.6	$   & 3.12& 3.13& 9.13&11.21 \\
\hline
\end{tabular}
\vspace{3mm}

\begin{minipage}{8.87cm}
\parindent=2.5mm
{\sc Notes}: Temperatures are in keV; luminosities are in units of
$10^{44}\,h^{-2}$\ergs; fluxes are corrected for absorption and in units of
$10^{-11}$\ergcms. Subscript $X$ denotes 0.1--2.4 keV energy band; bol
is for bolometric. Clusters with * are at $|b|<20^\circ$ and the flux of the
cluster with \dag\ is below the limit; these clusters are used for the
$L_X-T$ relation only.

$^{\rm a}$ Single-temperature fits and cooling flow-corrected
emission-weighted temperatures with 90\% errors are from M98, except those
marked d which are from David et al.\ (1993) and e which are estimated from
the $L_X-T$ relation derived here.

$^{\rm b}$ Total luminosities within $r<1\,h^{-1}$ Mpc (including cooling
flow regions).

$^{\rm c}$ Corrected quantities (cooling flow regions excised).
\end{minipage}
\end{center}
}
\end{minipage}
\vspace{4mm}

\begin{figure*}[tb]
\pspicture(0,-0.6)(18.5,8.6)

\rput[tl]{0}(0.,9.7){\epsfxsize=9cm
\epsffile{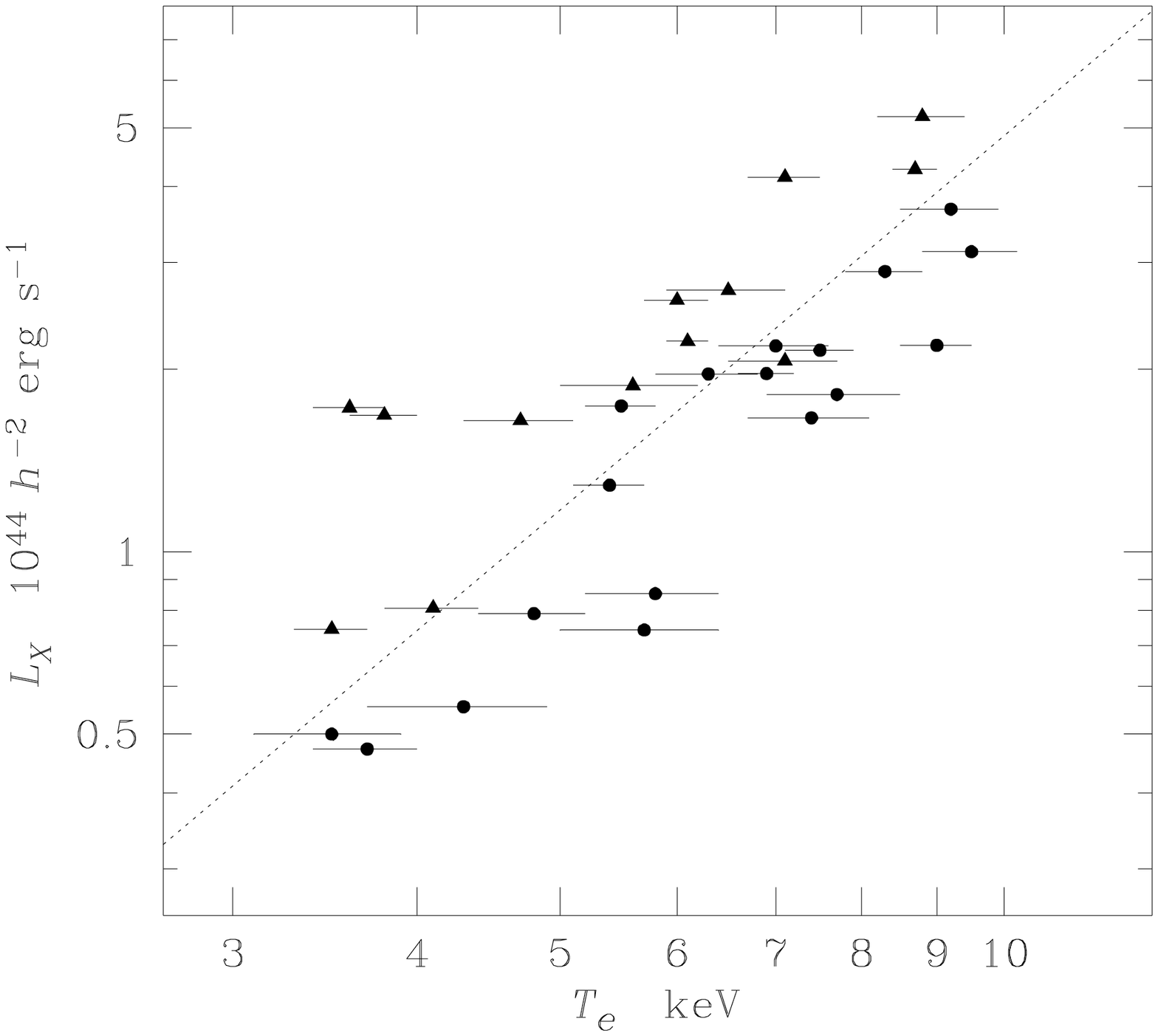}}

\rput[tl]{0}(9.6,9.7){\epsfxsize=9cm
\epsffile{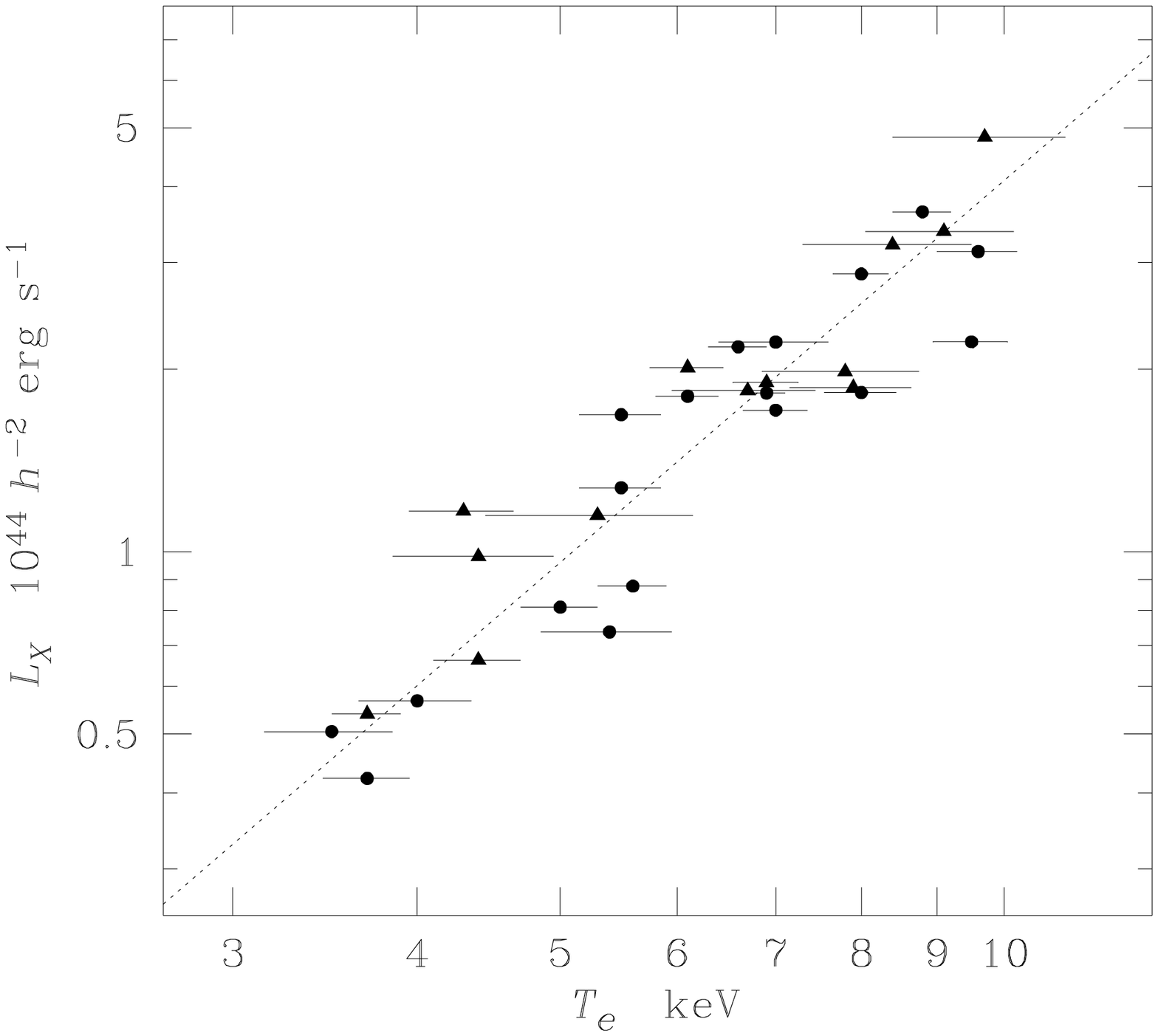}}

\rput[bl]{0}(1.8, 7.8){\it\large a}
\rput[bl]{0}(11.4,7.8){\it\large b}

\rput[tl]{0}(0,1.0){
\begin{minipage}{18.5cm}
\small\parindent=3.5mm
{\sc Fig.}~1.---Luminosities in the 0.1--2.4 keV band vs.\ temperatures from
Table 1. Panel ({\em a}) shows total luminosities within $r=1\,h^{-1}$ Mpc
and single-temperature fits to the overall \asca\ spectra, while panel ({\em
b}) shows corrected luminosities with cooling flow regions excised (see
text) vs.\ emission-weighted temperatures derived from the temperature maps
excluding cooling flow components. Triangles mark clusters with strong
cooling flows, circles are clusters with weak or no cooling flows. Dashed
lines are respective best-fit power laws from Table~2.
\end{minipage}
}
\endpspicture
\end{figure*}

For all clusters in the sample except 4, \asca\ wide-beam as well as
emission-weighted cooling flow-excluded temperatures are derived in M98 and
references therein. These temperatures correspond to averaging over
$r=0.6-1\,h^{-1}$ Mpc apertures, dependent on the individual observation
details; because the total cluster emission is dominated by bright central
regions, such a difference in radial coverage is unimportant. For the
remaining 4 clusters without the \asca\ data, temperatures are estimated
from the $L_X-T$ relation derived in \S\ref{lxt}. Non-\asca\ wide-beam
temperatures for these clusters exist (David et al.\ 1993) and are
consistent, within their relatively large uncertainties, with the estimated
values. We will use our estimates for the temperature function, since those
measured temperatures are not cooling flow-corrected.  For clusters with
\asca\ temperatures, Table~1 also lists cooling flow-excised bolometric
luminosities, estimated from the corrected 0.1--2.4 keV luminosities using
the corrected temperature and assuming isothermality. Our redshift
interval, imposed by the \asca\ analysis constraints (see M98), effectively
excludes clusters with $T\lax 3.5$ keV.

For the temperature function determination, as opposed to the $L_X-T$
relation, completeness of the sample is important. Therefore, as in previous
analyses, the $|b|<20^\circ$ area is excluded from the temperature function
subsample. Ebeling et al.\ (1996) find that their originally optically
selected list is already noticeably incomplete in the interval
$20^\circ<|b|<30^\circ$.  Excluding this area does not change our results,
so this small effect is ignored. Note that our sample, which consists of
bright, moderately nearby clusters with a flux limit a factor of 4 above the
completeness limit of the Ebeling et al.\ (1996) list, is less likely to be
affected by the above effect than their full list. In \S\ref{tfun}, we
confirm the completeness of our sample using the luminosity function from
the purely X-ray selected RASS cluster catalog (Ebeling et al.\ 1997; the
catalog itself is unavailable at the time of this writing).

The subsample of 30 clusters used for the temperature function has a median
temperature and redshift of about 6 keV and 0.054, respectively. It overlaps
to a large degree the samples of E90 and HA, both selected by 2--10 keV
flux, but differs from them mainly at low temperatures and fluxes. Unlike
our sample, the E90 dataset is not limited by redshift, which improves
statistics by including several hot, distant clusters, but has the
disadvantage of complicating the evolutionary studies. Our sample has a
lower effective flux limit than HA and therefore has more clusters in the
corresponding temperature range. The main improvement upon previous datasets
is, of course, the higher-quality imaging and temperature data, including
spatially resolved temperature measurements, available to us.

\section{THE $L_X-T$ RELATION}
\label{lxt}

Table~1 shows that for strong cooling flow clusters, the cooling
flow-corrected temperatures are significantly higher than the wide-beam
values, and that the small central regions contain a significant fraction of
the total X-ray luminosity. Below the luminosity-temperature relation in its
traditional meaning, using wide-beam temperatures and total luminosities, is
compared to the one that corresponds only to the main cluster gas, excluding
cooling flows in clusters that have ones.

\begin{center}
\pspicture(0,0.4)(8.8,8.6)

\rput[tl]{0}(0.,9.7){\epsfxsize=8.8cm
\epsffile{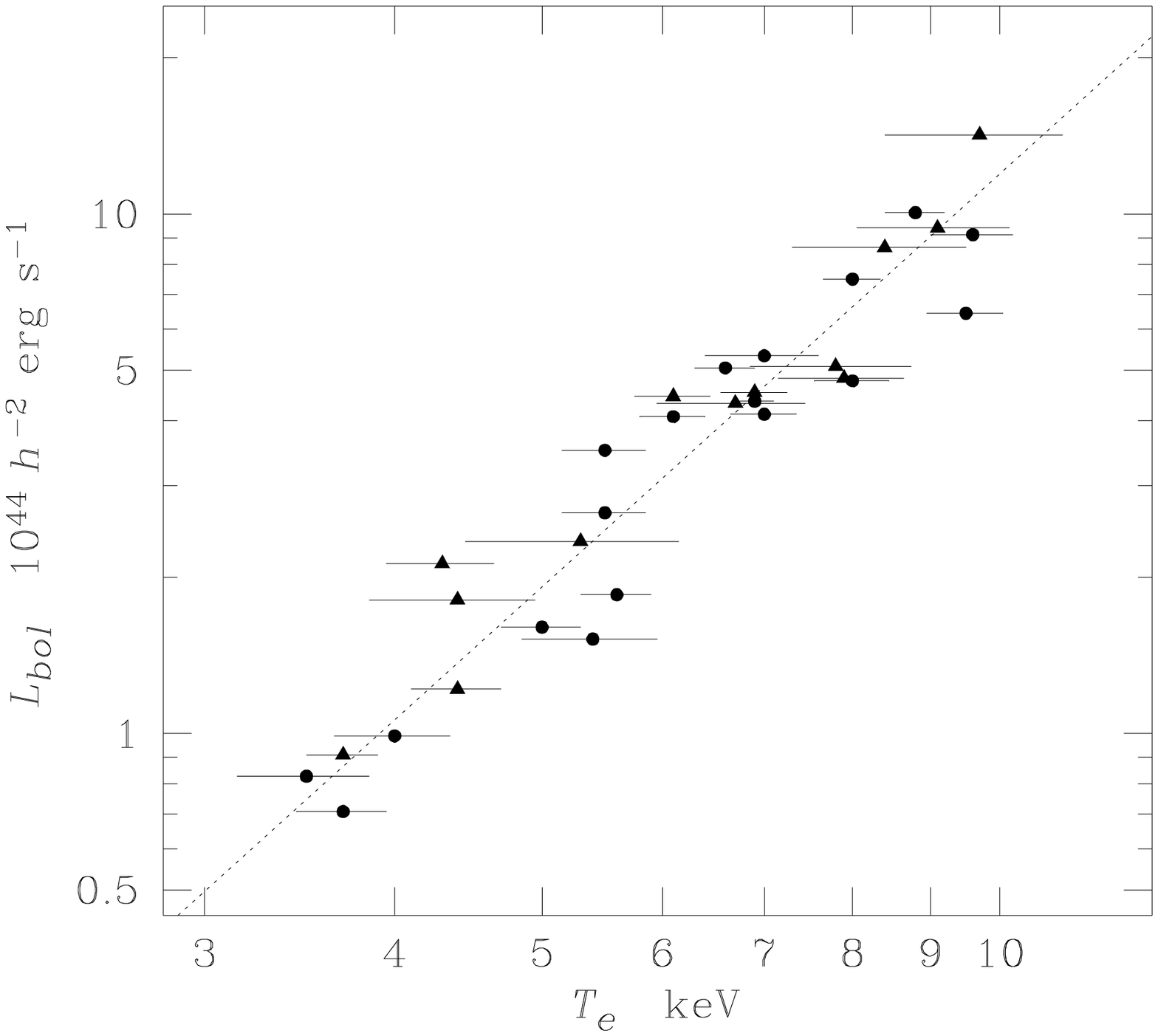}}

\rput[tl]{0}(0.,1.0){
\begin{minipage}{8.75cm}
\small\parindent=3.5mm
{\sc Fig.}~2.---Same as Fig.\ 1{\em b} (luminosities vs.\ temperatures
both corrected for cooling flows) but for bolometric luminosities.
\end{minipage}
}
\endpspicture
\end{center}

Figure~1 shows luminosities in the 0.1--2.4 keV band vs.\ temperatures for
clusters with \asca\ data, including and excluding the cooling flow regions.
Corrected bolometric luminosities vs.\ corrected temperatures are plotted in
Fig.\ 2. These $L-T$ relations are fit by power laws of the form
$L=A_6^{}T_6^\alpha$, where $T_6\equiv T/6\;{\rm keV}$, using the bisector
modification of the Akritas \& Bershady (1996 and references therein) linear
regression algorithm that allows for intrinsic scatter and nonuniform
measurement errors and treats $L$ and $T$ symmetrically. The confidence
intervals are evaluated by bootstrap resampling (e.g., Press et al.\ 1992)
while simultaneously adding random measurement errors to the temperatures
and 5\% errors to the luminosities (the results do not depend on the
latter). Table~2 lists the best-fit parameters together with the residual
rms scatter in $\log L$ for a given $T$, $\sigma_{\log L}$ (the scatter
along the $T$ axis is $\sigma_{\log T} \simeq \sigma_{\log L}/\alpha$). We
also include a fit for the cooling flow-corrected temperatures vs.\ total
luminosities, which may be useful, for example, to relate theoretical
predictions to the fluxes of poorly resolved distant clusters. Note that the
0.1--2.4 keV gas emissivity is only weakly dependent on temperature in the
considered temperature range, so the slope $\alpha$ of the $L_{\rm bol}-T$
relation should be greater than that for the 0.1--2.4 keV band by about 0.5
without any corrections.

{\footnotesize
\renewcommand{\arraystretch}{1.2}
\renewcommand{\tabcolsep}{1.5mm}
\begin{center}
TABLE 2
\vspace{1mm}

{\sc Fits to $L-T$ Relations}
\vspace{1mm}

\begin{tabular}{ccccc}
\hline \hline
Band & Data  & $A_6 $ & $\alpha$ & $\sigma_{\log L}$ \\
keV  &       & $10^{44}\,h^{-2}$\ergs & & \\
\hline
0.1--2.4   & Uncorr.\ $T$, total $L$& $1.71\pm0.21$ & $2.02\pm0.40$ & 0.181 \\ 
0.1--2.4   & Corr.\ $T$, total $L$ &  $1.57\pm0.17$ & $2.09\pm0.29$ & 0.133 \\
0.1--2.4   & Corr.\ $T$, corr.\ $L$&  $1.41\pm0.12$ & $2.10\pm0.24$ & 0.104 \\
Bol.       & Corr.\ $T$, corr.\ $L$&  $3.11\pm0.27$ & $2.64\pm0.27$ & 0.103 \\
\hline
\end{tabular}
\end{center}
}
\vspace{2mm}

Figure~1 and Table~2 show that the scatter in the $L_X-T$ relation is
greatly reduced when cooling flows are excised and the quantities
representing only the main cluster gas are compared. An example of the two
largest deviations around 3.7 keV in Fig.\ 1{\em a}, which are A780 (Hydra
A) and A2597, shows that the improvement is due in equal degrees to the
corrections of the temperature and luminosity. The same can be seen in the
last column of Table~2. Such a strong effect of the cooling flows on the
wide-beam temperatures was not anticipated before \asca, and the cooling
flow luminosity alone is insufficient to account for the scatter. This led
Fabian et al.\ (1994) to suggest that cooling flow clusters are different
from others in their global gas properties. Our results show that most of
the scatter in the $L_X-T$ relation is due to localized, $r\lax 50\,h^{-1}$
kpc regions with cooling flows. On the other hand, after the exclusion of
these regions, some residual scatter remains, which indeed is likely to be
due to the different gas distributions --- after the corrections, the
largest deviation at 9.5 keV in Fig.\ 1{\em b} is a strong merger A754. Note
that a small fraction of the scatter is due to the temperature measurement
errors.

The derived $L_X-T$ relation and its intrinsic scatter can be compared to
theoretical and numerical predictions that at present do not model radiative
cooling adequately. A self-similar model of cluster growth (Kaiser 1986),
supported by simulations that include only gravity (e.g., Navarro et al.\
1995), predicts $L_{\rm bol}\propto T^2$. A disagreement of this prediction
and the observed steeper slope has long been noted; our new $L_{\rm bol}-T$
relation is also significantly steeper. To explain this disagreement, it was
proposed that hypothetical preheating of the intracluster gas, e.g., by
supernovae, effectively reduces the central gas densities with a stronger
effect on the cooler clusters (e.g., Kaiser 1991; Evrard \& Henry 1991;
Navarro et al.\ 1995; Cavaliere et al.\ 1997). Assuming that, after initial
preheating, gas in the cluster cores remains on the same adiabat as clusters
evolve, Evrard \& Henry predict a scale-free relation $L_{\rm bol}\propto
T^{2.75}$. However, cluster merging should result in significant increase of
the gas specific entropy due to shock-heating, as is indeed observed (e.g.,
Markevitch et al.\ 1998b). Cavaliere et al.\ allow the gas inside the cores
to be shock-heated and mixed in addition to an early injection of the energy
equivalent to 0.5--0.8 keV per gas particle.  Their predicted $L_{\rm
bol}-T$ slope changes from 5 on the group scale to 2 on the hottest cluster
scale, being $\sim 2.3$ in the temperature interval covered by our data.
They also predict a scatter in the $L_{\rm bol}-T$ relation of about
$\sigma_{\log L}\simeq 0.13$ due to the differing merging histories of
individual clusters. Considering the number of adjustable quantities in such
modeling, these predictions are quite close to our observations. Although
our data are adequately described by a single power law, with such a small
scatter it is quite feasible to measure the predicted steepening of the
$L-T$ slope toward lower temperatures, if comparably accurate data at those
temperatures are obtained. Such a measurement would provide interesting
information on the energetics of the intracluster gas.

The observed tight $L_X-T$ correlation has other implications. M. Arnaud
(1997, private communication) used the small scatter of $L_X$ for clusters
lacking cooling flows to constrain the cluster-to-cluster baryon fraction
variations. We also note for illustration that our scatter in low-redshift
$L_X$ is equivalent to $0.25^m$, which is comparable to $\sim 0.19^m$ for
Type Ia supernovae used for cosmological distance estimates (Perlmutter et
al.\ 1997). However, clusters are not expected to remain standard candles at
all redshifts.

After this paper has been submitted for publication, we received a preprint
by Allen \& Fabian (1998) who employed a different approach to evaluate the
effect of cooling flows on the $L_X-T$ relation for hot clusters. They
modeled the overall cluster spectra by a sum of the isothermal and cooling
flow components and took the temperature of the isothermal component as the
true average cluster temperature. This compares to the M98 approach of using
spatially resolved spectra to localize cooling flow regions, obtain
temperature maps outside these regions, and use these maps to calculate
average temperatures. Allen \& Fabian find generally similar temperature
bias for those clusters studied in both works, although the exact
temperature values are sometimes different due to the difference in
approach. Unlike our result, the Allen \& Fabian's corrected $L_X-T$
relation still has considerable scatter and shows difference between
clusters with and without cooling flows. This is because their luminosities
include cooling flow regions, while we excise these regions when calculating
the corrected luminosities. Nevertheless, the slope of the corrected $L_{\rm
bol}\propto T^\alpha$ relation obtained by Allen \& Fabian,
$\alpha=2.3^{+0.6}_{-0.3}$, is in good agreement with our result.

\section{THE TEMPERATURE FUNCTION}

\subsection{Derivation}
\label{tfun}

The temperature function (TF), which is the number of clusters per unit
comoving volume and unit temperature interval, $dN/dT$, or the number above
a given temperature, $N(>T)$, can be derived by co-adding all clusters with
weights $1/V(T)$.  Here $V(T)$ is the maximum comoving volume within which a
cluster with a given temperature could have been detected above the flux
limit of our sample, and $T$ is a cooling flow-corrected temperature. Since
the sample is flux-limited rather than temperature-limited, an $L_X-T$
relation has to be used to calculate the volume as a function of
temperature. The sample was selected using the fluxes with excised cooling
flows and we can take advantage of the tight $L_X-T$ relation obtained
above. The volume as a function of temperature was calculated by
integrating, over the redshift interval 0.04--0.09, volume elements with
weights equal to the probability of a cluster with a given temperature to
have a detectable flux, assuming a log-Gaussian scatter in the $L_X-T$
relation given in Table~2.  For a purely flux-limited sample, a symmetric
$L_X-T$ scatter such as that results in a uniform increase of the volume by
a factor of about 1.06, compared to no scatter.\footnote{The reason is
geometric; the redshift shell $[z,z+\Delta z]$ containing the overluminous
clusters has greater volume than the shell $[z-\Delta z,z]$ containing the
underluminous ones.}
In the presence of the redshift constraints, the scatter results in the
volume correction between a factor of 1.26 at our minimum $T=3.5$ keV and
0.88 at 6.5 keV. The sample is essentially volume-limited above $\sim 7$
keV.

Note that this approach to the derivation of the TF differs from the one
employed, e.g., in HA and other works, in which individual cluster
luminosities (rather than temperatures) are used to determine the volume
within which a given cluster would have a detectable flux. The latter method
is entirely appropriate for the derivation of a luminosity function, but it
is not optimal for the TF, because clusters with a given temperature have a
wide range of luminosities. This method does result in an unbiased, on
average, TF estimate in the limit of the large number of clusters. In
reality, however, the number of clusters is always limited, and the large
intrinsic scatter in the $L_X-T$ relation causes bias in the median and most
probable TF estimates, especially at the highest (lowest) temperatures where
the TF is estimated using only one or two hottest (coolest) clusters in the
sample. To quantify this bias for a purely flux-limited sample, we performed
a simple simulation, scattering randomly in Euclidean space clusters with
luminosities distributed with an rms Gaussian scatter in $\log L_X$ of 0.28
(which corresponds to the $L_X-T$ scatter in HA). For those simulated
clusters above the flux limit, search volumes were calculated using
individual luminosities, $V\propto L^{3/2}$. The resulting distribution of
the calculated weights $1/V$ is strongly skewed; the median derived TF is
only 0.65 of the average (true) value at a temperature of the single hottest
cluster in the sample, while the most probable TF value is only 0.32 of the
average. The reason is that for any temperature, there are many more
overluminous clusters in a flux-limited sample than underluminous ones (for
the adopted value of the $L_X-T$ scatter, the ratio is 5:1).\footnote{In
principle, this may also affect the validity of the $L_X-T$ relation derived
from a flux-limited sample.  However, for our sample, selected by the
corrected fluxes and with a much smaller scatter in the corrected $L_X-T$
relation, this effect is small.}
The TF derivation method employed in this paper is free of this small-number
bias. Its possible disadvantage is the reliance on the validity of the
$L_X-T$ relation at all temperatures, including the extreme ones where the
number of clusters in the $L_X-T$ sample is small. However, this does not
apply to our sample, since it is volume-limited above $\sim 7$ keV, which
means the search volume for hot clusters is simply the total volume between
$0.04<z<0.09$.

The uncertainty of the resulting temperature function is evaluated by
bootstrap resampling, assuming a Poissonian distribution of the total number
of clusters in the sample and adding random Gaussian measurement errors to
the individual temperatures at each resampling. The latter is a minor
contribution to the overall uncertainty dominated by that of the small
number of clusters.

The resulting cumulative temperature function is plotted in Fig.\ 3,
overlaid in panel ({\em a}) on the previously derived functions of E90 and
HA (the latter is taken from Eke et al.\ 1996). The figure shows that the
overall effect of our corrections for the presence of cooling flows at each
stage of the derivation is not large --- there is agreement with the earlier
E90 and HA functions within their $1\sigma$ uncertainties (not shown). The
cooling flow correction on average results in a small shift of the TF toward
higher temperatures. An apparent (not highly significant) disagreement
between our and HA results at high temperatures is mostly due to (a) the
fact that our sample is volume-limited while the HA sample is flux-limited,
(b) a possible incompleteness of the HA sample (e.g., it does not include
the hot cluster A2163 which appears to satisfy the selection criteria), and
(c) the possibility of the small-number bias described above.

To assess the completeness of our sample, particularly at low temperatures
where the Abell catalog may miss objects even at high Galactic latitudes
(e.g., MKW3S), we use the luminosity function obtained from the purely X-ray
selected RASS cluster catalog (Ebeling et al.\ 1997) and convert it into the
temperature function using the $L_X-T$ relation and its scatter. The
relevant $L_X-T$ relation was derived using the total, RASS-measured
(Ebeling et al.\ 1996) luminosities and corrected temperatures of our
clusters. The resulting relation is similar to line 2 in Table~2 but has a
larger scatter due to the RASS flux uncertainties. The resulting temperature
function, shown in Fig.\ 3{\em a}, is in excellent agreement with our
result, indicating that no significant number of cool clusters is missing.
The luminosity function was derived from a much larger cluster sample than
ours so its uncertainty is negligible.

To obtain an analytic fit to the differential temperature function, we use
fine temperature binning and maximum likelihood minimization assuming a
Poissonian distribution of the cluster number in each bin (bins are allowed
to have zero clusters; the results are independent of the bin size). The
errors are calculated by the bootstrap algorithm outlined above. For a power
law of the form $dN/dT=A_6^{}T_6^{-\alpha}$, we obtain, assuming
$\Omega_0=1$, $A_6=1.90\pm0.54\times 10^{-7}\,h^3\,{\rm Mpc}^{-3}\,{\rm
keV}^{-1}$ and $\alpha=4.2\pm0.7$. A slightly better fit is obtained for an
exponential $A_6\exp[-(T-6\;{\rm keV})/T_*]$ with $A_6=2.38\pm0.68\times
10^{-7}\,h^3\,{\rm Mpc}^{-3}\,{\rm keV}^{-1}$ and $T_*=1.6\pm0.4$ keV.
These models are shown in cumulative form in Fig.\ 3{\em b}.

\begin{figure*}[thb]
\pspicture(0,-0.5)(18.5,8.6)

\rput[tl]{0}(0.,9.7){\epsfxsize=9cm
\epsffile{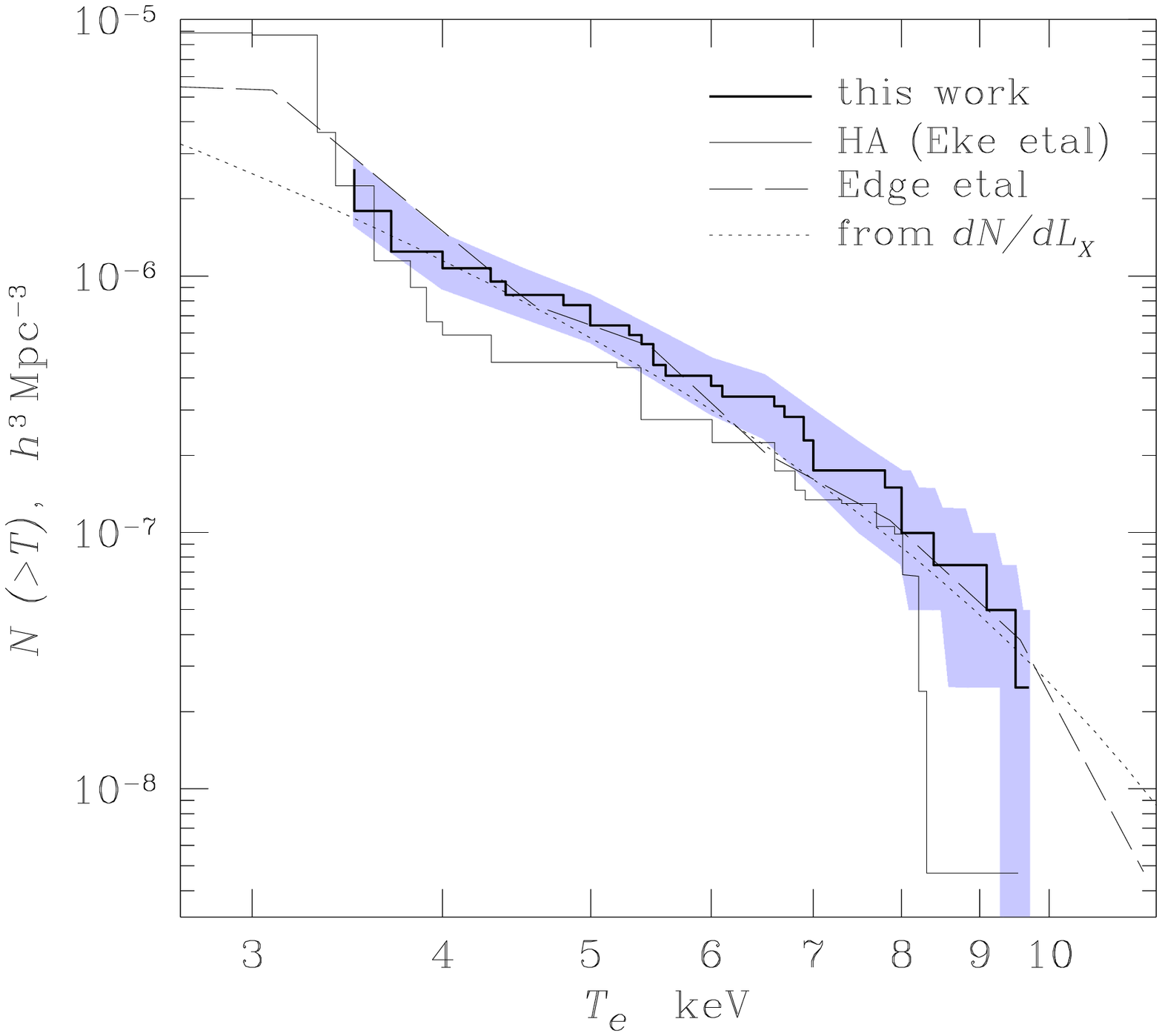}}

\rput[tl]{0}(9.6,9.7){\epsfxsize=9cm
\epsffile{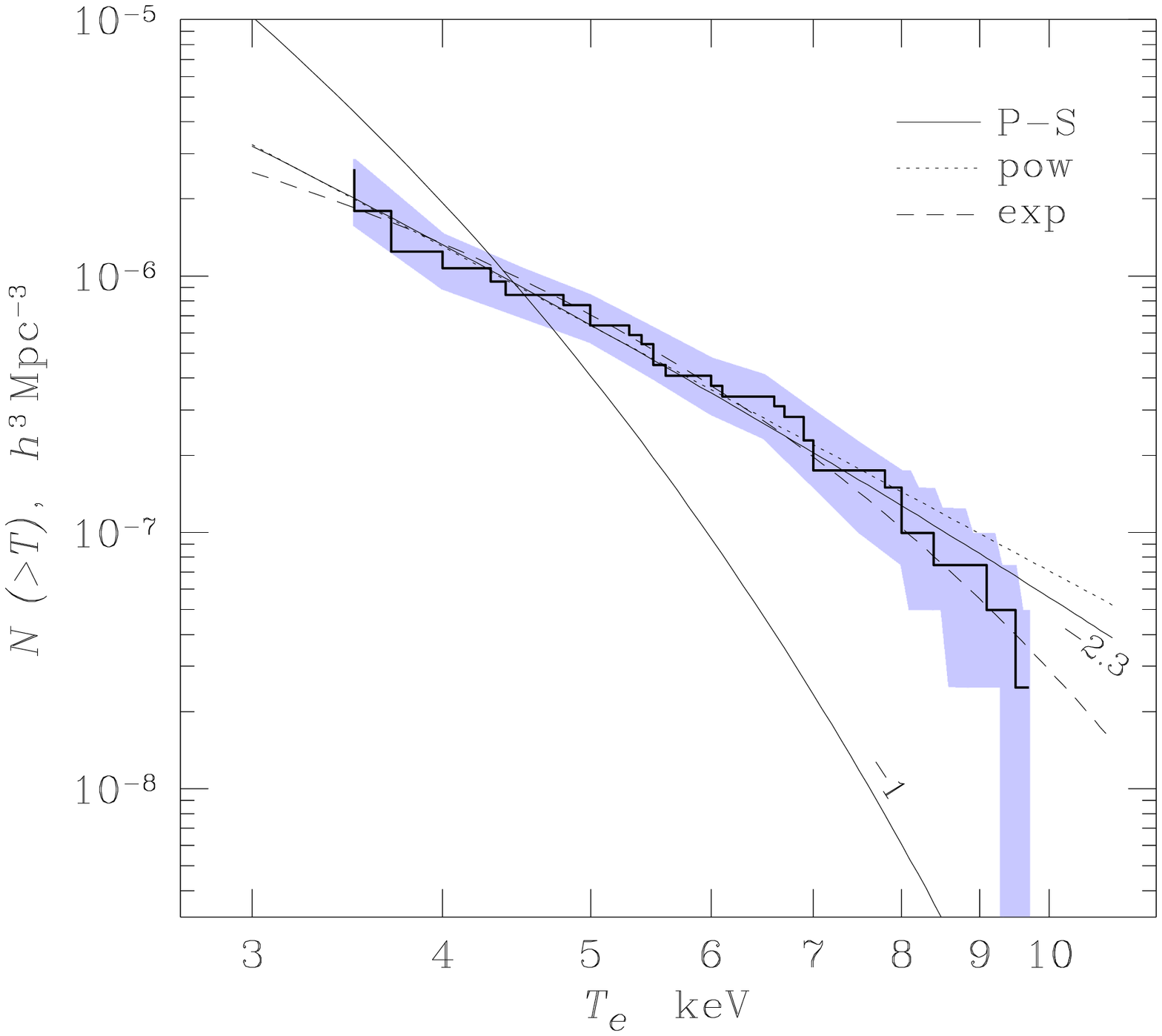}}

\rput[bl]{0}(1.8, 2.5){\it\large a}
\rput[bl]{0}(11.4,2.5){\it\large b}

\rput[tl]{0}(0,1.0){
\begin{minipage}{18.5cm}
\small\parindent=3.5mm
{\sc Fig.}~3.---Cumulative temperature function. Panel ({\em a}) shows our
result (assuming $\Omega_0=1$) with its 68\% error band overlaid on the
earlier derivations of HA and E90, as well as a function obtained from the
luminosity function for a purely X-ray selected sample (Ebeling et al.\
1997) using the $L_X-T$ relation (see text). Panel ({\em b}) shows fits to
our data: Press-Schechter fits for $n=-1$ (standard CDM) and $n=-2.3$
(best-fit for $\Omega_0=1$), and empirical power law and exponential fits (see
text).
\end{minipage}
}
\endpspicture
\end{figure*}

\subsection{Constraining the Fluctuation Spectrum}

The cluster temperature function was used to obtain constraints on the
cosmological density fluctuation spectrum by many authors (the incomplete
listing includes HA; Kaiser 1991; White, Efstathiou, \& Frenk 1993; Viana \&
Liddle 1996; Eke et al.\ 1996; Colafrancesco, Mazzotta, \& Vittorio 1997;
Pen 1998) applying the Press \& Schechter (1974) formalism supplemented by
simulations. Our temperature function is slightly higher and flatter than
the previous estimates, and its accuracy is better. In addition, Markevitch
\& Vikhlinin (1997) and M98 recently showed that the observed cluster
nonisothermality considerably reduces the X-ray mass estimate for a given
emission-weighted temperature, compared to the previously made estimates. An
accurate comparison of cosmological models to the data requires hydrodynamic
simulations to take into account the nonequilibrium state of many clusters.
However, we can perform a qualitative estimate of the effect of the above
observational updates on the fluctuation spectrum constraints by applying
the Press-Schechter formalism for $\Omega_0=1$ and $\Omega_0=0.3$
($\Lambda=0$), and comparing the results to similar previous analyses. A
full description of the formalism can be found in the references above, so
only the relevant definitions are given here.

We assume that the power spectrum of linear density fluctuations on the
cluster scale is $P_k\propto k^n$, implying an rms amplitude of the linear
mass fluctuations in a sphere containing an average mass $M$ of
$\sigma(M)=\sigma_8\, (M/m_8)^{-(n+3)/6} (1+z)^{-1}$, where $m_8$ is the
average mass within the comoving $r=8\,h^{-1}$ Mpc.  Our clusters are
assumed to lie at the median sample redshift of 0.054, being observed just
after their collapse (a reasonable assumption for $\Omega_0=1$ but not
necessarily such for $\Omega_0\ll 1$, e.g., White \& Rees 1978;
nevertheless, we use this assumption for $\Omega_0=0.3$ for comparison with
other works). The cluster mass within a radius of overdensity $\Delta_c$
(taking $\Delta_c=178$ for $\Omega_0=1$ and $\Delta_c=124$ for
$\Omega_0=0.3$ as in Eke et al.\ 1996), is assumed to scale with the X-ray
emission-weighted temperature as $M=M_{10}(T/10\;{\rm keV})^{3/2}$ with an
rms scatter of 20\% as predicted by simulations of Evrard, Metzler, \&
Navarro (1996). 

It is worth mentioning here that the scaling $M\propto T^{3/2}$ follows from
the same line of argument as the relation $L_{\rm bol}\propto T^2$, and the
latter is not observed. Processes that modify the $L_X-T$ relation, e.g.,
preheating, can possibly break the $M-T$ scaling. However, X-ray luminosity
is very sensitive to the gas density distribution, which is more likely to
change due to preheating than the gas temperature. Indeed, simulations by
Metzler \& Evrard (1998) suggest that the $M-T$ relation holds at least for
their considered preheating model, while the gas density distributions (and
therefore the $L_X-T$ relation) are significantly modified. The available
observational data (masses from gravitational lensing vs.\ X-ray
temperatures) at least do not contradict the adopted $M-T$ relation (Hjorth,
Oukbir, \& van Kampen 1998). X-ray measurements of masses at different
temperatures (our work in progress) will further test the accuracy of this
scaling.

For $M_{10}$ we use two values for each assumed $\Omega_0$ (values in
parentheses correspond to $\Omega_0=0.3$), $13.5\;(16.4) \times
10^{14}\,h^{-1}$\msun\ which assumes cluster isothermality (e.g., Eke et
al.\ 1996) and $9.1\;(9.8) \times 10^{14}\,h^{-1}$\msun\ which corresponds
to the mass profile of the main cluster of A2256 derived by Markevitch \&
Vikhlinin using the observed temperature profile. Although the latter values
are based on a single cluster, temperature profiles for many other clusters
are similar and the effect on the mass is expected to be similar as well
(M98). Using these relations, $m_8=6\times 10^{14}\,\Omega_0 h^{-1}$\msun\
corresponds to $T\simeq 3-8$ keV for $\Omega_0=0.3-1$, indicating that the
cluster data best constrain the density fluctuations at this particular
scale.

The Press-Schechter mass function was converted to the differential
temperature function which was then fitted to the data as described in
\S\ref{tfun}, with $\sigma_8$ and $n$ being free parameters. The observed TF
was derived self-consistently using the two values of $\Omega_0$.  The
best-fit $\sigma_8$ and $n$ are given in Table~3 for the two cosmologies and
two assumed cluster temperature distributions. The best-fit model TF in the
cumulative form for $\Omega_0=1$ is shown in Fig.\ 3{\em b} (the two
temperature profile cases are almost identical). As was noted in previous
works, because the abundance of clusters is very sensitive to $\sigma_8$,
the latter quantity is rather robust to the uncertainty in the number of
clusters. Therefore, even though our temperature function is about a factor
of 2 above that of HA used in most previous works, our $\sigma_8$ for the
isothermal case and $\Omega_0=1$ is only slightly higher than the values
derived from the corrected HA data (0.50, Eke et al.\ 1996, 1998; 0.53, Pen
1998). Our result for $\Omega_0=0.3$ is significantly different from the
above two works which report $\sigma_8\simeq 0.9$ for $\Omega_0=0.3$. The
difference is due to our treatment of $n$ as a free parameter; fixing
$n\simeq -1.3$ or $-1.4$ as in those works and assuming isothermality
results in best-fit $\sigma_8$ values close to those in the above works, for
both values of $\Omega_0$. However, such values of $n$ do not fit the data.

\vspace{1mm}
{\footnotesize
\renewcommand{\arraystretch}{1.2}
\renewcommand{\tabcolsep}{1.5mm}
\begin{center}
TABLE 3
\vspace{1mm}

{\sc Press-Schechter Fits to the Temperature Function}
\vspace{1mm}

\begin{tabular}{lccccc}
\hline \hline
$T$ {\sc profile} & \multicolumn{2}{c}{$\Omega_0=1$} & &
                  \multicolumn{2}{c}{$\Omega_0=0.3$} \\
\cline{2-3} \cline{5-6} 
            & $\sigma_8$ & $n$ & & $\sigma_8$ & $n$ \\
\hline
{\sc Isothermal} &$0.55\pm0.03$ &$-2.3\pm0.3$ &&$0.78\pm0.04$ & $-2.0\pm0.3$ \\
{\sc Observed}   &$0.51\pm0.03$ &$-2.3\pm0.3$ &&$0.66\pm0.03$ & $-2.1\pm0.3$ \\
\hline
\end{tabular}
\end{center}
}

\vspace{2mm}
\begin{center}
\pspicture(0,-3.8)(8.8,8.6)

\rput[tl]{0}(0.,9.7){\epsfxsize=8.8cm
\epsffile{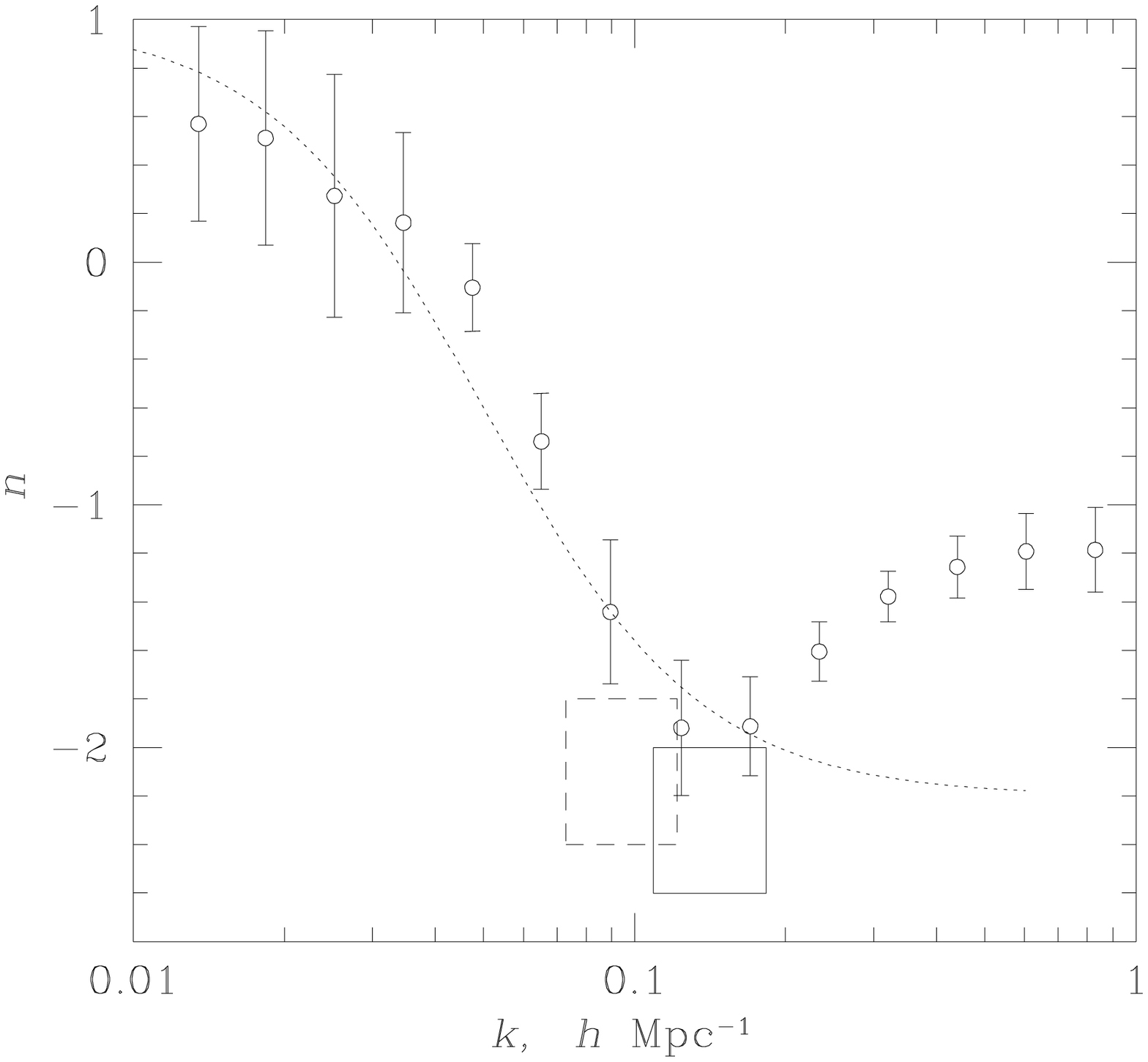}}

\rput[tl]{0}(0,1.0){
\begin{minipage}{8.75cm}
\small\parindent=3.5mm
{\sc Fig.}~4.---Circles with $1\sigma$ errors show the slope of the APM
galaxy density fluctuation spectrum ($P_k\propto k^n$) as a function of
scale, from Baugh \& Gazta\~naga (1996). Flattening of the observed spectrum
at $k\gax 0.2$ reflects the onset of nonlinear fluctuation growth; dotted
line is a fit to the de-evolved, linear spectrum. Rectangles shows our slope
of the linear spectrum (solid line for $\Omega_0=1$, dashed line for
$\Omega_0=0.3$). Their vertical sizes are the 90\% errors, while their
horizontal sizes illustrate the approximate $k$ coverage corresponding to
our temperature range. The rectangles are plotted at $k=1/r$, where $r$ is
the radius containing the average mass $M$ corresponding to a given
temperature. A more meaningful average $k$ weighted by the contribution of
each harmonic to $\sigma(M)$ depends on the exact spectrum, so the rectangle
positions are rather approximate.
\end{minipage}
}
\endpspicture
\end{center}

The derived $\sigma_8$ is sensitive to the assumed mass-temperature
relation. The observed nonisothermality, which modifies this relation, acts
to reduce the derived value (see Table~3); the scatter in the relation,
included in all our estimates, also reduces the $\sigma_8$ estimate by $\sim
0.01$. Since some of the observed radial temperature decline may be due to
the incomplete thermalization of the gas in the outermost cluster regions
(biasing low the X-ray-derived total mass), detailed cluster simulations
reproducing such a decline are necessary for an accurate interpretation of
the temperature function. In the absence of such simulations, the $\sigma_8$
values obtained above may be viewed as conservative brackets of the true
value for the assumed cosmologies.

The best-fit fluctuation spectrum slope is $n=-2.3$ and $n\simeq -2$ for
$\Omega_0=1$ and $\Omega_0=0.3$, respectively; the results are very close
for isothermal and nonisothermal cases (see Table~3). The value
corresponding to $\Omega_0=1$ is consistent with $n\simeq -2$ obtained by
Kaiser (1991) using the E90 data and is marginally steeper than
$n=-(1.7^{+0.65}_{-0.35})$ (68\% two-parameter interval) reported by HA and
$n=-1.8\pm0.4$ by Colafrancesco et al.\ (1997) using the HA data. Our weak
dependence of the best-fit $n$ values on $\Omega_0$ seems to disagree with
the Colafrancesco et al.\ finding; however, it is difficult to compare fits
to the TFs with rather different observed shapes.

Our temperature interval corresponds to the log range in the wavenumber $k$
of only 0.23 (although a comparable interval of $k$ contributes to the mass
fluctuation $\sigma(M)$ at any $M$), thus it probes a local shape of the
fluctuation spectrum. Coincidentally, the galaxy fluctuation spectrum
derived from the APM survey, which is the most accurate determination to
date (Baugh \& Efstathiou 1993 and references therein; Baugh \& Gazta\~naga
1996), exhibits maximum steepness at this scale with a local slope of
$n\simeq -1.9\pm0.3$ (68\% interval). Figure 4 shows the observed APM slope
as a function of scale from Baugh \& Gazta\~naga together with our
measurement.  Flattening at $k\gax 0.2$ reflects the onset of nonlinear
fluctuation growth; the above authors recover the approximate primordial,
linear spectrum (dotted line in Fig.\ 4) with $n\simeq -2$ on these scales.
The reconstructed linear-regime slope as well as the observed slope are in
agreement with our X-ray results.  It is worth recalling that the galaxy
distribution is linked to the underlying, almost linear, density field via
an unknown and possibly scale-dependent bias. Clusters, on the other hand,
represent nonlinearly evolved density peaks and their abundance constrains
the density spectrum directly, under the assumption of gaussianity of the
fluctuations (and, of course, assuming that the X-ray-measured cluster
masses are not grossly in error). Agreement between two such widely
different methods is encouraging.  Both the optical and the X-ray slope are
considerably steeper than the standard CDM prediction of $n\simeq -1$ at
this scale (e.g., Blumenthal et al.\ 1984).  For illustration of the
discrepancy with the X-ray data, Fig.\ 3{\em b} shows a temperature function
for $n=-1$. A discussion of some promising alternatives (namely, mixed dark
matter) can be found, e.g., in Baugh \& Efstathiou (1993).

\section{SUMMARY AND FUTURE WORK}

Using the new \asca\ cluster temperatures and \rosat\ luminosities, both
calculated directly excluding the cooling flow regions, we obtain the
$L_X-T$ relation in the 3.5--10 keV temperature interval with greatly
reduced scatter. This result provides an accurate reference point ({\em a})
for a comparison with nearby, cooler clusters and groups that should have a
different slope if there was preheating, and ({\em b}) for evolution studies
using the oncoming \axaf\ data which will provide the necessary angular
resolution for cooling flow excision at higher redshifts. The derived
$L_X-T$ relation can be directly compared to theoretical and numerical
predictions, most of which do not include a detailed treatment of radiative
cooling.

These new data are also used to rederive the nearby cluster temperature
function. This function is interpreted qualitatively by applying the
Press-Schechter formalism for $\Omega_0=1$ and $\Omega_0=0.3$, to obtain
values of $\sigma_8$ assuming the isothermal and observed cluster
temperature profiles (see Table~3). These values should bracket the correct
value if some of the observed radial temperature decline in clusters is due
to the incomplete thermalization of the gas in the outer regions. The
derived slope of the mass fluctuation spectrum at the cluster scale is
$n=-(2.0-2.3)\pm0.3$ for considered values of $\Omega_0$, which is
consistent with the APM galaxy survey at the same scale. Again, our
temperature function provides a reference for future accurate evolution
studies and for direct comparison with theoretical predictions.

It would be interesting to extend the temperature range of this work to
lower temperatures, 1--3 keV, where one expects to see, for example, the
effects of preheating on the $L_X-T$ relation. It would also greatly
strengthen the constraints from the temperature function. Such an extension
will be possible when X-ray selected cluster catalogs become available.
Since preheating can modify the simple mass-temperature scaling, accurate
measurements of cluster masses over a range of temperatures, including the
lowest, will be required for the interpretation of such an extended
temperature function.

\acknowledgments

The author is grateful to Alexey Vikhlinin, Bill Forman and Vincent Eke for
useful discussions and comments. The results reported here would not be
possible without the dedicated work of the \asca\ and \rosat\ teams on
building, calibration and operation of the instruments. The HEASARC online
data archive at NASA/GSFC has been used extensively in this research.  The
work was supported by NASA contract NAS8-39073.

\end{document}